\documentclass[12pt]{article}
\usepackage{amssymb,amsmath,epsfig}

\begin{document}

\title{\textbf{Thermodynamics of a Bardeen black hole in noncommutative space}}

\author{M. Sharif \thanks{msharif.math@pu.edu.pk} and Wajiha Javed \\
\\Department of Mathematics, University of the Punjab,\\
Quaid-e-Azam Campus Lahore-54590, Pakistan.}
\date{}

\maketitle
\begin{abstract}
In this paper, we examine the effects of space noncommutativity on
the thermodynamics of a Bardeen charged regular black hole. For a
suitable choice of sets of parameters, the behavior of the
singularity, horizon, mass function, black hole mass, temperature,
entropy and its differential, area and energy distribution of the
Bardeen solution have been discussed graphically for both
noncommutative and commutative spaces. Graphs show that the
commutative coordinates extrapolate all such quantities (except
temperature) for a given set of parameters. It is interesting to
mention here that these sets of parameters provide the singularity
(essential for $r_h>0$) and horizon ($f(r_h)=0$ for $r_h>0$) for
the black hole solution in noncommutative space, while for
commutative space no such quantity exists.
\end{abstract}
{\bf Keywords:} Regular black hole; Noncommutative space;
Thermodynamics.\\
{\bf PACS:} 04.70.Dy; 04.70.Bw; 11.25.-w

\section{Introduction}

It is well known that the classical theory of relativity breaks
down at Planck's scale, $l_p$, where quantum gravitational effects
become important. To solve this problem, it is convenient to alter
Riemannian structure of space-time by some mathematical model. As
in relativistic physics, gravity distorts the geometry of
space-time; similarly, quantum gravity quantizes space-time.
According to von Neumann, the geometrical properties of quantum
mechanics are described as "pointless structure", which is known
as "noncommutative geometry". Noncommutative geometry helps to
handle space-time structure at very short distances \cite{B1}.

Space-time noncommutativity is based on quantum mechanics. The
replacement of canonical position $x^i$ and momentum variables
$p_j$ with corresponding Hermitian operators,
$\hat{x}^i,~\hat{p}_j$, defines the quantum phase space that obeys
Heisenberg commutation relations
\begin{equation}
[\hat{x}^i,\hat{p}_j]=\iota\hbar\delta^i_j\quad i,j=1,2,3\label{1}
\end{equation}

The limit $\hbar\longrightarrow 0$ recovers an ordinary space. A
noncommutative space can be realized by the coordinate operators,
$\hat{x}_i$, satisfying the commutation relations
\begin{equation}
[\hat{x}^i,\hat{x}^j]=\iota\theta^{ij}\label{2}
\end{equation}
where the noncommutative parameter $\theta^{ij}$ is a constant,
real valued, antisymmetric tensor, and has dimensions of
(length)$^2$ \cite{B2}. The tensor $\theta^{ij}$ is generally
assumed to depend on space-time coordinates and momenta.

The space-time quantization arises in the following two problems.
The first is the renormalization problem, and the second is
quantum gravity. The model of quantum space-time must satisfy the
following canonical commutation relations in noncommutative space
(with $\hbar=1$) \cite{B1}:
\begin{equation}
[\hat{x}_i,\hat{x}_j]=\iota\theta_{ij}\quad[\hat{x}_i,\hat{p}_j]=
\iota\delta_{ij}\quad[\hat{p}_i,\hat{p}_j]=0\label{3}
\end{equation}
with
\begin{equation}
\theta_{ij}\theta^{ij}=0\quad
\epsilon_{ijkl}\theta^{ij}\theta^{kl}=-8l^4_p\label{2}
\end{equation}

The replacement of the usual multiplication of fields with the
$\star$-product provides the formulation of noncommutative field
theories from commutative field theories \cite{B3}. The
$\star$-product is defined in terms of $\theta_{ij}$ as follows
\cite{B4}:
\begin{equation}
(f\star
g)(x)=\exp\left(\frac{\iota}{2}\theta^{ij}\frac{\partial}{\partial
x^i}\frac{\partial}{\partial y^j}\right)f(x)g(y)|_{x=y}\label{999}
\end{equation}
where $f$ and $g$ are infinitely differentiable arbitrary functions.

Myung and Yoon \cite{B11} found a three-dimensional new regular
black hole with two horizons by introducing an anisotropic perfect
fluid inspired by a four-dimensional noncommutative black hole.
They compared the thermodynamics of this black hole with that of a
nonrotating Ba$\tilde{n}$ados-Teitelboim-Zanelli (BTZ) black hole.
Tejeiro and Larranaga \cite{B12} investigated a new rotating
regular black hole and analyzed its thermodynamics. The results
were compared with those of a rotating BTZ solution. Nicolini et
al. \cite{B13} studied the effects of noncommutativity on the
terminal phase of the Schwarzschild black hole evaporation. Nozari
and Fazlpour \cite{B14,B15} explored the behavior of
noncommutative space and the generalized uncertainty principle on
the thermodynamics of a radiating Schwarzschild black hole. They
also considered the effects of space noncommutativity on the
thermodynamics of a Reissner-Nordstr$\ddot{o}$m (RN) black hole.
Nozari and Mehdipour \cite{B18} explored space noncommutativity by
generalizing the RN black hole in dD space-time and also discussed
its thermodynamics. Ansoldi et al. \cite{B16} found a
noncommutative geometry inspired solution of the Einstein-Maxwell
field equations describing a variety of charged self-gravitating
objects. Spallucci et al. \cite{B17} obtained a noncommutative
solution to the field equations in higher dimensions.

Nasseri \cite{B5} investigated the effects of noncommutative
spaces on the horizon, the area spectrum, and Hawking temperature
of the Schwarzschild black hole. Sadeghi \cite{B6} studied
noncommutative spaces in a two-dimensional black hole. He obtained
the event horizon in noncommutative space up to second-order
perturbation and a lower limit for the noncommutativity parameter.
Also, Sadeghi and Setare \cite{B7} performed the noncommutative
corrections to the behavior of a BTZ black hole. Alavi \cite{B4}
investigated the parameters of a noncommutative RN black hole. He
also studied the stability of the black hole and found an upper
bound for the noncommutativity parameter, $\theta$.

This paper is devoted to investigating the behavior of the Bardeen
charged regular black hole solution in noncommutative space. The
thermodynamical quantities in this noncommutative space are
studied. We compute quantities like singularity, horizon, mass
function, black hole mass, Hawking temperature, entropy and its
differential, area and energy distribution in noncommutative space
and compare their behaviors in commutative and noncommutative
spaces graphically.

In Sect. 2, we propose the usual Bardeen black hole in
noncommutative space. Also, we analyze graphically the effects of
space noncommutativity and commutativity on the singularity and
horizon of a given regular black hole. Section 3 describes
graphical behavior of space noncommutativity and commutativity of
the thermodynamical quantities. Finally, we summarize the results
in the last section.

\section{Space noncommutativity}

The Bardeen model \cite{B19} was proposed some years ago as a
regular black hole obeying the weak energy condition. The metric
for the Bardeen model (with $G=1$) is given by
\begin{equation}
{ds}^2=\left(1-\frac{2M}{r}\right){dt}^2-\frac{{dr}^2}
{\left(1-\frac{2M}{r}\right)}-{r^2}({d\theta}^2+\sin^2\theta{d\phi}^2)\label{}
\end{equation}
where
\begin{equation}
M=M(r)=\frac{mr^3}{(r^2+e^2)^\frac{3}{2}}\label{ghi}
\end{equation}
is a mass function. This solution is a self-gravitating magnetic
monopole with charge $e$ of nonlinear electrodynamic source and
exhibits black hole behavior for $e^2\leq 16m^2/27$. For $e=0$,
the solution reduces to the Schwarzschild space-time.

The spherical event horizons, $r_h$, are given by the roots of the
equation ($g_{tt}=0$)
\begin{equation*}
r_h=2M(r_h)\label{}
\end{equation*}

Substituting the value of $M$, we get
\begin{equation}
1-\frac{2mr^2_h}{(r_h^2+e^2)^\frac{3}{2}}=0\label{lmno}
\end{equation}

We introduce the following metric for the Bardeen black hole in
noncommutative space \cite{B5}:
\begin{equation}
{ds}^2=\left(1-\frac{2M}{\sqrt{\hat{r}\hat{r}}}\right){dt}^2-
\frac{d\hat{r}d\hat{r}}{\left(1-\frac{2M}
{\sqrt{\hat{r}\hat{r}}}\right)}-{\hat{r}\hat{r}}({d\theta}^2+\sin^2\theta{d\phi}^2)\label{4}
\end{equation}
where $\hat{r}$ satisfies (\ref{3}). The event horizon,
$\hat{r}_h$, of the space-time (\ref{4}) in noncommutative space
satisfies the following condition:
\begin{equation}
1-\frac{2m\hat{r}_h\hat{r}_h}{(\hat{r}_h\hat{r}_h+e^2)^{\frac{3}{2}}}=0\label{4a}
\end{equation}

The $\star$-product (\ref{999}) between two fields in
noncommutative phase space can be replaced by a shift, called
Bopp's shift, and can be defined as \cite{N}
\begin{equation}
x_i=\hat{x}_i+\frac{1}{2}\theta_{ij}\hat{p}_j\quad
p_i=\hat{p}_i\label{5}
\end{equation}
where $x_i$ and $p_i$ satisfy the usual (commutative) commutation
algebra
\begin{equation}
[x_i,x_j]=0\quad [x_i,p_j]=i\delta_{ij}\quad[p_i,p_j]=0\label{}
\end{equation}

Thus the effects of space-space noncommutativity can be calculated
in commutative space. When we apply transformation (\ref{5}) to
line element (\ref{4}), this turns out to be in terms of the
position and momentum variables of commutative phase space. Using
coordinate transformation (\ref{5}) from $\hat{r}_h$ to $r_i$,
(\ref{4a}) takes the following form:
\begin{equation}
1-\frac{2m(r_i-\frac{\theta_{ij}p_j}{2\hbar})(r_i-\frac{\theta_{ik}p_k}
{2\hbar})}{\left[(r_i-\frac{\theta_{ij}p_j}{2\hbar})(r_i-\frac{\theta_{ik}p_k}
{2\hbar})+e^2\right]^\frac{3}{2}}=0\label{}
\end{equation}

This leads to the relations
\begin{equation}
1-\frac{2m}{r}-\frac{1}{r^3}\left({mr_i\theta_{ij}p_j}-{3me^2}\right)
-\frac{3me^2r_i\theta_{ij}p_j}{r^5}+\ldots=0\label{}
\end{equation}
or
\begin{equation}
1-\frac{2m}{r}-\frac{1}{r^3}\left(\frac{m\textbf{L}.\theta}{2}-3me^2\right)
-\frac{3me^2\textbf{L}.\theta}{2r^5}+\ldots=0\label{6}
\end{equation}
where $\theta_{ij}=(1/2)\epsilon_{ijk}\theta_k$ and
$\textbf{L}=\textbf{r}\times \textbf{p}$ is the angular momentum.
If we set $\theta_3=\theta$ and assume that all the remaining
components of $\theta$ vanish (which can be done by rotation or
redefinition of the coordinates), then
$\textbf{L}.\theta=L_z\theta$. In this case, (\ref{6}) takes the
form
\begin{equation}
r^5-2mr^4-r^2\left(\frac{mL_z\theta}{2}-3me^2\right)-\frac{3me^2L_z\theta}{2}=0\label{}
\end{equation}
\begin{center}
{\bf {\small Table 1.}} {\small For a singularity, the behavior of
$f(r_h)$ with sets of parameters.}

\vspace{0.25cm}

\begin{tabular}{|l|l|l|l|l|}
\hline { Color} & {\bf $m$}&{\bf $L_z\theta$}&{\bf $e$}
\\ \hline {Orange}  & $.00098755$&$.00001$&500.5
\\ \hline {Blue} & $.55$&$-1$&.12
\\ \hline {Purple} & $.01$&$-150$&.15
\\ \hline {Grey} & $.1$&$-.05$&1
\\ \hline
\end{tabular}
\end{center}
\begin{figure}
\center\epsfig{file=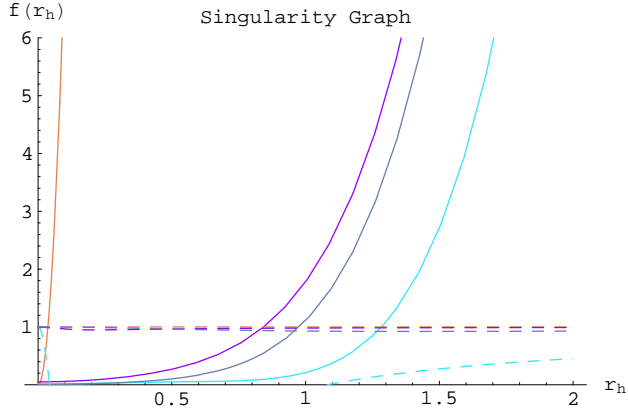, width=0.65\linewidth}
\caption{Horizon function, $f(r_h)$, versus horizon radius, $r_h$:
the solid lines represent singularity in noncommutative space for
different sets of parameters given in Table 1. The dashed lines do
not represent singularity in commutative space for the same sets
of parameters.}
\end{figure}

In Fig. 1, it is clear that $f(r_h)\longrightarrow\infty$ for
$r_h>0$ with the sets of parameters given in Table 1 (for solid
lines). This is because the noncommutative parameter $\theta$ is
nonzero in this case. The dashed lines imply that
$f(r_h)\longrightarrow1$ for $r_h=0$ with the same sets of
parameters. This shows that for the commutative case there is no
singularity and curves extrapolate the distance function.
\begin{center}
{\bf {\small Table 2.}} {\small For the horizon, the behavior of
$f(r_h)$ with sets of parameters.}

\vspace{0.25cm}

\begin{tabular}{|l|l|l|l|l|}
\hline { Color}  & {\bf $m$}&{\bf $L_z\theta$}&{\bf $e$}
\\ \hline { Red}  & $.1$&$-.009$&.9
\\ \hline { Blue} & $.0002$&$1$&3.5
\\ \hline { Green} & $.125$&$1.15$&1.5
\\ \hline { Black} & $.1$&$.0005$&.2
\\ \hline
\end{tabular}
\end{center}
\begin{figure}
\center\epsfig{file=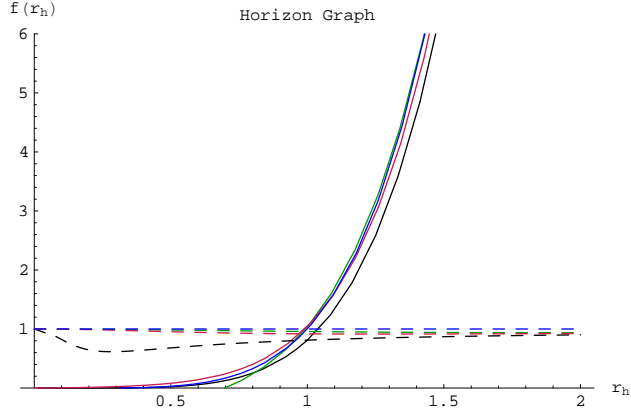, width=0.65\linewidth}
\caption{Horizon function, $f(r_h)$, versus horizon radius, $r_h$:
the four solid curves represent the horizon for the Bardeen black
hole in noncommutative space with different sets of parameters
given in Table 2. The dashed lines do not represent the horizon
for the Bardeen black hole in commutative space with the same sets
of parameters.}
\end{figure}

In Fig. 2, all four solid curves represent horizons corresponding
to Bardeen solution (\ref{4}) in noncommutative space with the
choice of parameters given in Table 2. The dashed curves for the
commutative case with the same sets of parameters indicate that
there is no curve for which $f(r_h)=0$ with nonzero horizon
radius, $r_h\neq0$. Here, distances also extrapolate. Thus, each
of these curves does not have a meaning similar to $r=2M$ with
$M=mr^3/(r^2+e^2)^{3/2}$ in the case of the usual Bardeen
solution.

\section{Thermodynamics}

This section is devoted to investigating graphically the behavior
of thermodynamical quantities in noncommutative space. For this
purpose, we choose sets of parameters given in Table 2 with
horizon radii that provide four curves representing a function of
horizon radius. We investigate the effects of noncommutativity on
the thermodynamical quantities (mass function, black hole mass,
temperature, entropy and its differential, area and energy
distribution). The mass function for the Bardeen model in
commutative space is given by (\ref{ghi}), while for $\hbar=1$,
its form in noncommutative space can be written as
\begin{equation}
\hat{M}=m\left(1-\frac{3e^2}{2r_h^2}+\frac{9e^2{L_z\theta}}{8r_h^4}\right)\label{}
\end{equation}

The mass of the Bardeen regular black hole (\ref{lmno}) is given
by
\begin{equation}
m=\frac{(r_h^2+e^2)^{\frac{3}{2}}}{2{r_h^2}}\label{}
\end{equation}
while in noncommutative space, it turns out to be
\begin{equation}
\hat{m}=\frac{r_h}{2}\left(1-\frac{{L_z\theta}}{4r_h^2}+
\frac{3e^2}{2r_h^2}+\frac{3e^2{L_z\theta}}{4r_h^4}\right)\label{}
\end{equation}
\begin{figure}
\center\epsfig{file=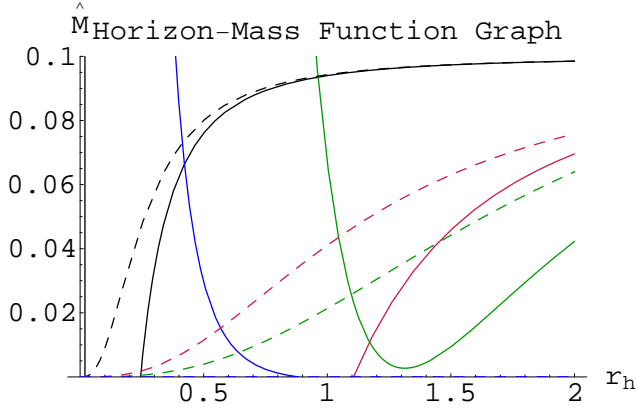, width=0.65\linewidth} \caption{Mass
function, $\hat{M}$, versus horizon radius, $r_h$: the solid lines
represent behavior of the mass function, $\hat{M}$, in
noncommutative space for different sets of parameters given in
Table 2. The dashed lines represent its behavior in commutative
space for the same sets of parameters.}
\end{figure}
\begin{figure}
\center\epsfig{file=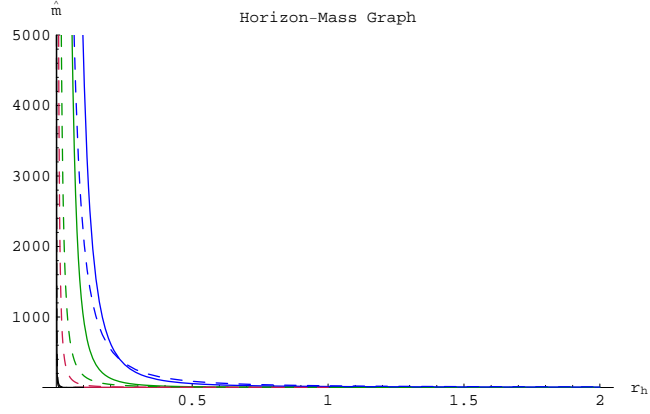, width=0.65\linewidth}
\caption{Black hole mass $\hat{m}$ versus horizon radius $r_h$:
the solid lines represent behavior of the black hole mass
$\hat{m}$ in noncommutative space while the dashed lines represent
its behavior in commutative space.}
\end{figure}

In Fig. 3, the solid curves give the representation of the mass
function $\hat{M}(r_h)$ in noncommutative space. The black curve
shows asymptotic and increasing behavior whereas blue and red
represent decreasing and increasing behavior, respectively, while
green shows both of them. The dashed lines correspond to the mass
function, $M(r_h)$, in commutative space. Here, the blue curve
represents zero mass function while the remaining curves show
increasing behavior. Curves in commutative space extrapolate the
mass function.

Figure 4 presents the graphical behavior of the black hole mass.
This shows that black, blue and green curves have the same
behavior in both commutative and noncommutative spaces. These
curves show asymptotic and decreasing behavior while the red curve
shows similar behavior in both spaces. It is noteworthy that, for
commutative space, the black hole mass always exhibits asymptotic
behavior in the region where $m>0$. Noncommutative coordinates
show the extrapolation for the black hole mass.

The usual Bardeen black hole temperature, $T_H$, is given by
\cite{B8}
\begin{equation}
T_H={\frac{\hbar \partial_rF(r)}{4\pi}}|_{r=r_h}=\frac{\hbar m
r_h(r_h^2-2 e^2)}{2\pi(r_h^2+e^2)^{\frac{5}{2}}} \label{7}
\end{equation}
where $F(r)=[1-(2M/r)]$ and $m=(r^2+e^2)^{3/2}/2r^2$. For
$\hbar=1$, we obtain the following Hawking temperature in
noncommutative space:
\begin{figure}
\center\epsfig{file=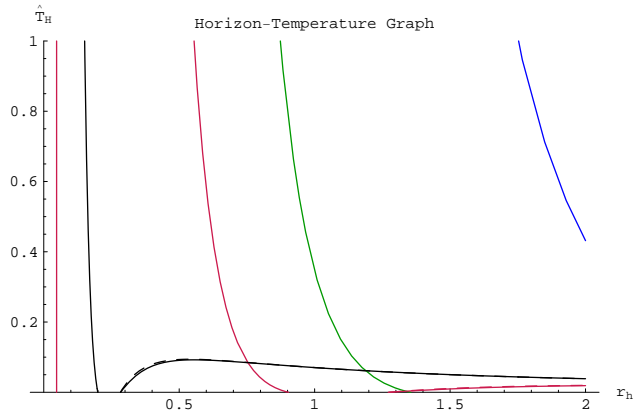, width=0.65\linewidth}
\caption{Hawking temperature, $\hat{T}_H$, versus horizon radius,
$r_h$: the solid lines indicate the behavior of temperature,
$\hat{T}_H$, in noncommutative space while the dashed lines
represent its behavior in commutative space.}
\end{figure}
\begin{equation}
\hat{T}_H=\frac{1}{4\pi r_h}\left(1+\frac{L_z\theta}{4r_h^2}
-\frac{3e^2}{r_h^2}-\frac{5{e^2}{L_z\theta}}{4r_h^4}
+\frac{2e^4}{r_h^4}+\frac{{e^4}{L_z\theta}}{2r_h^6}\right)\label{}
\end{equation}

It is mentioned here that for large black holes (i.e.,
$r_h^2/4\theta\gg 0$) we recover the temperature of the Bardeen
regular black hole (\ref{7}).

Each curve in Fig. 5 shows the behavior of temperature. For
noncommutative space, horizon radius and temperature, $\hat{T}_H$,
of a Bardeen black hole preserve the inverse relation for solid
green and blue curves while for the red and black it indicates
asymptotic, decreasing, and increasing behavior. For commutative
space, dashed green and blue curves indicate that the behavior of
temperature, $T_H$, does not lie in the region of positive
temperature, $T_H>0$. However, red and black curves have
overlapping behavior with noncommutative space and indicate that
$T_H\rightarrow0$ with increasing $r_h$. Thus, for all $T_H$,
commutative coordinates do not show extrapolation for a given set
of parameters.

The first law of thermodynamics for a Bardeen regular black hole is
\begin{equation}
dm=TdS+\phi de\label{}
\end{equation}
where the electric potential of the black hole is given by
\begin{equation}
\phi=\frac{\partial m}{\partial e}|_{r=r_h}=\frac{3 e}{2
r_h^2}(r_h^2+e^2)^\frac{1}{2}\label{}
\end{equation}

The entropy of the black hole is \cite{B8}
\begin{equation}
S={\int_{r_0}^{r_h}}\frac{1}{T_H}dm=2\pi\hbar^{-1}\left((-\frac{e^2}{r_h}+
\frac{r_h}{2})\sqrt{e^2+r_h^2}+
\frac{3}{2}{e^2}\ln(r_h+\sqrt{e^2+r_h^2})\right)\label{abc}
\end{equation}
\begin{figure}
\center\epsfig{file=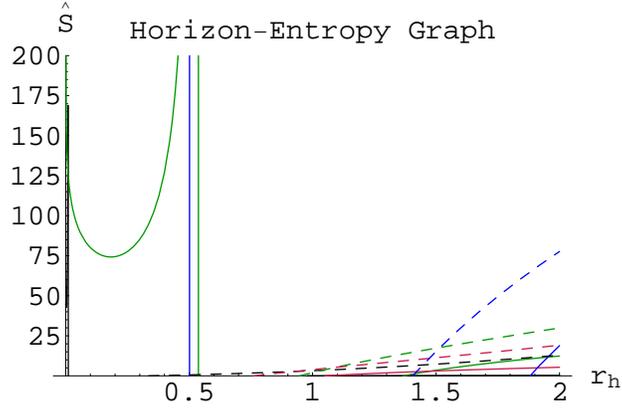, width=0.65\linewidth}
\caption{Entropy, $\hat{S}$, versus horizon radius, $r_h$: the
solid lines represent entropy, $\hat{S}$, of the black hole in
noncommutative space while the dashed lines represent the entropy,
$S$, of the black hole in commutative space.}
\end{figure}
\begin{figure}
\center\epsfig{file=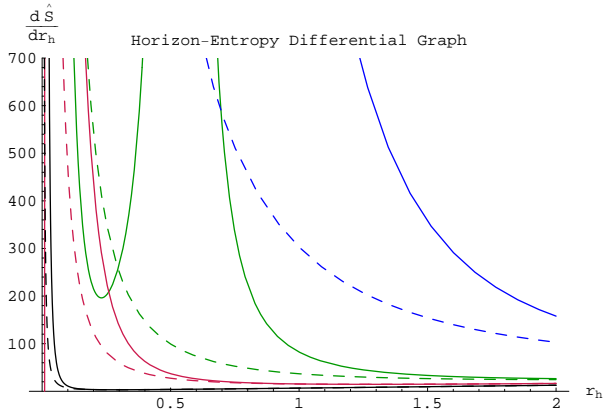, width=0.65\linewidth}
\caption{Derivative of entropy with respect to horizon radius,
$\frac{d\hat{S}}{dr_h}$, versus horizon radius, $r_h$: the solid
lines show the behavior of $\frac{d\hat{S}}{dr_h}$ in
noncommutative space while the dashed lines show it in commutative
space.}
\end{figure}

In noncommutative space, it becomes
\begin{eqnarray}
\hat{S}&=&2\pi[r_h(-\frac{4{e^2}r_h}{4r_h^2
-L_z\theta}+\frac{r_h}{2}-\frac{L_z\theta}{8r_h})
(1+\frac{1}{2r_h^2}(e^2-\frac{L_z\theta}{2}))\nonumber\\&+&\frac{3}{2}
e^2\ln(r_h-\frac{L_z\theta}{4r_h}+r_h(1+
\frac{1}{2r_h^2}(e^2-\frac{L_z\theta}{2})))]\label{def}
\end{eqnarray}

The entropy as a function of $r_h$ is represented in Fig. 6. In
the large black hole limit, the entropy function corresponds to
the Bekenstein-Hawking area law relating entropy and horizon area
(i.e., $S=A/4$) for the regular black hole geometry.

The derivative of entropy, $S$, (\ref{abc}) with respect to
horizon radius, $r_h$, in commutative space can be written as
\begin{eqnarray}
\frac{dS}{dr_h}&=&\frac{2\pi}{\hbar}[(-\frac{e^2}{r_h}
+\frac{r_h}{2})\frac{r_h}{\sqrt{e^2+r_h^2}}
+\sqrt{e^2+r_h^2}(\frac{e^2}{r_h^2}+\frac{1}{2})\nonumber\\
&+&\frac{3e^2}{2(r_h+\sqrt{e^2+r_h^2)}}(1+\frac{r_h}{\sqrt{e^2+r_h^2}})]
\end{eqnarray}
In noncommutative space ($\hbar=1$), it becomes
\begin{eqnarray}
\frac{d\hat{S}}{dr_h}&=&2\pi[(\frac{-4e^2r_h}{4r_h^2-L_z\theta}
+\frac{r_h}{2}-\frac{L_z\theta}{8r_h})(1-\frac{1}{2r_h^2}(e^2-\frac{L_z\theta}{2}))\nonumber\\
&+&(r_h+\frac{1}{2r_h}(e^2-\frac{L_z\theta}{2}))(\frac{4e^2(L_z\theta
+4r_h^2)}{(4r_h^2-L_z\theta)^2}+\frac{1}{2}+\frac{L_z\theta}{8r_h^2})\nonumber\\
&+&\frac{3e^2(4r_h^2+L_z\theta-e^2)}{2r_h(4r_h^2-L_z\theta+e^2)}]
\end{eqnarray}

In Fig. 6, there are two different behaviors of the entropy,
$\hat{S}$, in noncommutative space represented by solid lines.
Black and red curves indicate asymptotic and linearly increasing
behavior, respectively, while blue and green show both of them.
The entropy, $S$, in commutative space is shown by dashed lines,
where all four of the curves exhibit linearly increasing behavior
while only red and black curves extrapolate black hole entropy.

In Fig. 7, both solid lines in noncommutative space indicate
entropy differential $\frac{d\hat{S}}{dr_h}$ and the dashed lines
in the commutative case represent $\frac{dS}{dr_h}$ and both
exhibit the same behavior. Green, red and black curves show
asymptotic behavior while the blue curve indicates that the
entropy differential decreases as horizon radius increases. All
four of the curves in commutative space extrapolate entropy
differential.

The horizon area, $A$, of the Bardeen regular black hole is given
by \cite{B5}
\begin{equation}
A=r^2_h{\int^{2\pi}_0}d\phi{\int^\pi_0}\sin\theta
d\theta=4\pi{r^2_h}=16\pi
M^2=\frac{16\pi{m^2}{r_h^6}}{(r_h^2+e^2)^3}\label{}
\end{equation}

For $\hbar=1$, this turns out to be the area in noncommutative
space
\begin{equation}
\hat{A}=\frac{16\pi m^2(r_h^6-\frac{3}{2}r_h^4
L_z\theta)}{(r_h^6-\frac{3}{2}L_z\theta r_h^4+3 e^2
r_h^4)}\label{}
\end{equation}
\begin{figure}
\center\epsfig{file=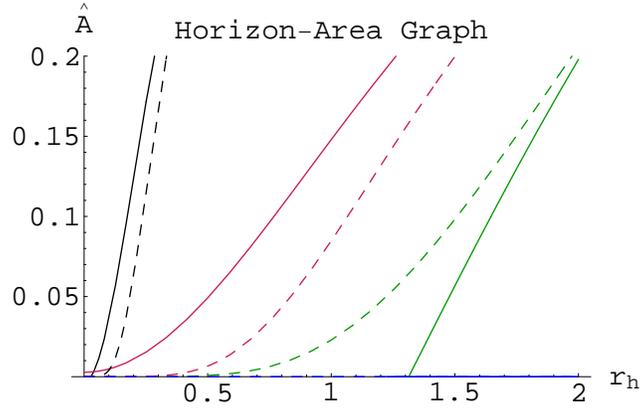, width=0.65\linewidth}
\caption{Horizon area, $\hat{A}$, versus horizon radius, $r_h$:
the solid lines show the area, $\hat{A}$, of the regular black
hole in noncommutative space while the dashed lines show the area
$A$ in commutative space.}
\end{figure}
\begin{figure}
\center\epsfig{file=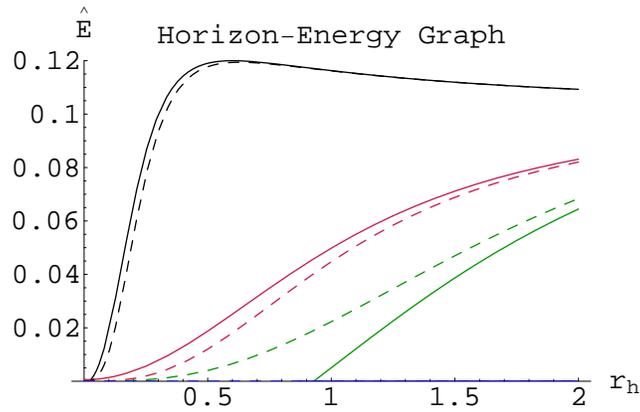, width=0.65\linewidth}
\caption{Energy distribution function, $\hat{E}$, versus horizon
radius, $r_h$: the solid lines represent the energy distribution
function, $\hat{E}$, in noncommutative space while the dashed
lines represent it in commutative space.}
\end{figure}

In Fig. 8, the solid lines demonstrate that the area, $\hat{A}$,
is zero, corresponding to the blue curve in noncommutative space.
The remaining three curves representing area increase linearly
with an increase in horizon radius. For commutative space, the
dashed lines display the same behavior of area, $A$, as the
noncommutative case. For commutative space, horizon area must be
positive and extrapolates for the given choice of parameters.

In the Landau-Lifshitz prescription, the energy distribution is
\cite{B9}
\begin{equation}
E=-\frac{mr_h^3}{-(r_h^2+e^2)^\frac{3}{2}+2mr_h^2}\label{}
\end{equation}

At large distances or for $e=0$, this reduces to the energy of the
Schwarzschild black hole \cite{B10}. In noncommutative space, it
becomes
\begin{equation}
\hat{E}=-\frac{m(r_h-\frac{3L_z\theta}{4r_h})}{r_h(1+\frac{3}{2r_h^2}
(e^2-\frac{L_z\theta}{2}))+2m(1-\frac{L_z\theta}{2r_h^2})}\label{}
\end{equation}
In Fig. 9, curves describe the same behavior as that for the
corresponding area graph for both noncommutative and commutative
spaces. The energy distribution function, $E$, of the regular
black hole in commutative space is always positive. Blue and Green
curves extrapolate energy in commutative space while black and red
have the same behavior in both spaces.

\section{Summary}

This paper deals with the effects of noncommutative space on the
behavior of singularity, horizon, and thermodynamical quantities
of Bardeen regular black holes. It also extends our recent work on
the thermodynamics of the usual Bardeen black hole \cite{B8}. For
a suitable choice of parameters, singularity, horizon, mass
function, black hole mass, temperature, entropy and its
differential, area and energy distribution have been investigated
and compared graphically in both commutative and noncommutative
spaces.

It has been found that, for horizon radius $r_h>0$, there is a
class of parameters for which function of horizon radius,
$f(r_h)$, approaches infinity, which leads to a singularity. If
the noncommutative parameter $\theta$ approaches zero, then for
$r_h=0$, $f(r_h)$ does not approach infinity, which leads to the
usual black hole solution without having the essential
singularity. For noncommutative space, the Bardeen solution
represents the horizon, while the Bardeen solution in commutative
space does not represent the horizon for the same sets of
parameters.

The behavior of the thermodynamical quantities can be summarized
as follows:
\begin{itemize}
\item ¢ Different curves of the mass function in noncommutative space
show decreasing, increasing, and asymptotic behavior while the
mass function has increasing behavior for $M>0$ in the commutative
case.
\item ¢ The black hole mass indicates asymptotic behavior for $m>0$ in
commutative geometry, while for
the noncommutative case, some curves exhibit asymptotic while some
represent similar behavior to the commutative case.
\item ¢ For the noncommutative case, temperature expresses asymptotic,
decreasing, and increasing behaviors. For the commutative case,
two curves have overlapping behavior with the noncommutative case
while temperature approaches zero and as horizon radius increases.
The remaining two curves do not show any behavior in the region
where $T_H\geq0$ but have asymptotic behavior in the region where
$T_H<0$.
\item ¢ The entropy has overlapping behavior (asymptotic and increasing)
in the noncommutative case, while it has only increasing behavior
in the commutative case. The entropy differential exhibits the
same behavior in both spaces.
\item ¢ Energy and area exhibit the same behavior in both spaces.
\item ¢ Commutative coordinates extrapolate distances and thermodynamic
quantities (except temperature) for a particular choice of
parameters.
\end{itemize}

It would be interesting to extend this analysis to examine the
behavior of thermodynamic quantities in noncommutative space for
the stringy charged black hole solutions and also for the regular
black hole solution with cosmological constant.

\vspace{0.25cm}

{\bf Acknowledgment}

\vspace{0.25cm}

We would like to thank the Higher Education Commission, Islamabad,
Pakistan, for its financial support through the Indigenous Ph.D.
5000 Fellowship Program Batch-IV.

\end{document}